\documentclass[11pt]{article}
\usepackage{amssymb}

\newcommand{\half}{\mbox{$\frac{1}{2}$}}
\newcommand{\fslash}{\!\!\not\!}
\newcommand{\dslash}{\!\!\not\!\partial}
\newcommand{\tr}{\mbox{tr}}
\newcommand{\Tr}{\mbox{Tr}}

\begin{document}

\begin{table}
\begin{flushright}
IK--TUW 9906401
\end{flushright}
\end{table}

\title{Chiral Poincar\'{e} transformations and their
anomalies\thanks{Supported by Fonds zur F\"{o}rderung der
wissenschaftlichen Forschung, P11387--PHY}}

\author{Jan B. Thomassen\thanks{E-mail: {\tt
thomasse@kph.tuwien.ac.at}} \\
{\em Institut f\"{u}r Kernphysik, Technische Universit\"{a}t Wien} \\
{\em A--1040 Vienna, Austria}}

\date{August 31, 1999}

\maketitle

\begin{abstract}

I consider global transformations of a Dirac fermion field, that are
generated by the generators of Poincar\'{e} transformations, but with
a $\gamma_5$ appended. Such chiral translations and chiral Lorentz
transformations are usually not symmetries of the Lagrangian, but
naively they are symmetries of the fermionic measure. However, by
using proper time regularization in Minkowski space, I find that they
in general give rise to a nontrivial Jacobian. In this sense they have
``anomalies''. I calculate these anomalies in a theory of a massive
fermion coupled to an external Abelian vector field. My motivation for
considering chiral Poincar\'{e} transformations is the possibility
that they are relevant to bosonization in four dimensions.

\vspace{\baselineskip}
\noindent
PACS numbers: 11.30.Rd; 11.15.Tk; 11.30.Qc \\
{\em Keywords}: Chiral anomalies; Bosonization; Proper time
regularization

\end{abstract}

\section{Introduction}

The global symmetries of the Lagrangian of a Dirac fermion include
phase transformations, spacetime translations and Lorentz
transformations. These symmetries are linked to charge, mass, and spin
-- quantum numbers that together characterize the particle states
described by the Lagrangian.

If a $\gamma_5$ is appended to the generators of these
transformations, we get chiral phase and chiral Poincar\'{e}
transformations. Although chiral phase transformations have been
extensively studied, little is known about chiral translations and
chiral Lorentz transformations. The obvious reason for this is that
they are not symmetries of the Lagrangian unless the fermion is free
and massless. However, chiral phase transformations are also not
symmetries of the Lagrangian if the fermion is massive, and still
these play a role in physics. In QCD, chiral phase transformations in
flavor space are believed to be approximate symmetries that are
spontaneously broken and give rise to a Goldstone boson octet with
$\pi$'s, $K$'s and the $\eta$ -- even if the quarks are massive. Thus
we cannot rule out the possibility that a transformation has physical
meaning on the grounds that it is not a symmetry.

Chiral phase rotations have the property that they are, naively,
symmetries of the fermionic measure. This is also true for the chiral
Poincar\'{e} transformations, and they are therefore singled out from
the set of all possible transformations of a Dirac fermion that are
not symmetries of the Lagrangian.

In this paper I investigate chiral Poincar\'{e} transformations within
a simple Abelian model of a fermion coupled to an external vector
field. My motivation for this has to do with bosonization
\cite{coleman}. While bosonization is a well established concept in
two dimensions (see e.g.\ \cite{damgaard} and references therein), the
same thing cannot be said about four dimensions (however, see
\cite{burgess}). If we consider the various bosonization schemes in
the path integral \cite{damgaard}, they all have the property that the
bosonic field in the bosonized theory is the chiral phase of the
fermion, and the bosonic action $S_{\mathrm{bos}}$ is given by the
Jacobian $J=\exp(iS_{\mathrm{bos}})$ of a local chiral transformation
of the fermion. But in four dimensions it cannot be expected that just
the chiral phase of a fermion is enough to describe the physics. It is
therefore necessary to identify further degrees of freedom of the
fermion that are relevant for bosonization. I think the degrees of
freedom that correspond to chiral Poincar\'{e} transformations are
good candidates for this.

If chiral Poincar\'{e} transformations really is to play a similar role
in four dimensional bosonization as chiral phase rotations does in the
two-dimensional case, then they must give rise to a Jacobian when a
change of variables is performed in the path integral. In this paper I
will restrict my attention to this Jacobian and calculate it for
infinitesimal chiral Lorentz transformations and translations. A
calculation of the Jacobian for {\em finite} transformations and an
investigation of the bosonization procedure itself will be reported in
a separate publication \cite{thomassen}.

The Jacobian $J[\beta]=\exp(i\int d^4x\beta{\cal A})$ of a chiral
rotation with parameter $\beta(x)$ contains the chiral anomaly ${\cal
A}(x)$. It is a quantum correction to the naive expectation that $J$
is unity. Thus, since the same thing happens for chiral Poincar\'{e}
transformations they also have a kind of ``anomaly''. This is a
distinguishing property of chiral Poincar\'{e} transformations and is
logically independent of the fact that they are not symmetries of the
Lagrangian. In lack of a better word, I will sometimes refer to the
quantity that corresponds to ${\cal A}$ for chiral Poincar\'{e}
transformations as `anomaly' as well.

I should also mention a paper by Alvarez \cite{alvarez} which is
slightly related to this work. He tried to calculate the fermionic
determinant of a fermion that is coupled to a vector field by using a
decoupling transformation. His decoupling transformation corresponds
to the spin part of a chiral Lorentz transformation. There is also the
recent paper \cite{cortes} where a similar decoupling transformation
is used. However, my work is not directly related to investigations of
{\em local} Lorentz symmetry \cite{witten}.

The organization of the paper is the following. In sec.\ 2, I discuss
classical properties of the chiral Poincar\'{e} transformations. I
point out that they make sense if they are regarded as active,
transforming the physical system, rather than passive.  In sec.\ 3, I
consider the quantum theory which I regularize using the proper time
representation of the fermionic determinant in Minkowski space. I
prefer this to the usual Euclidean formulation (see e.g.\ the review
by Ball \cite{ball}), because better control over the anomalies is
achieved.  In sec.\ 4, I calculate the Jacobians, which is a
generalization of the calculation that leads to the
Adler--Bell--Jackiw (ABJ) anomaly \cite{fujikawa}. The new features
which we must give attention to is the presence of derivative
operators and a $\sigma_{\mu\nu}$-matrix in the generators. In sec.\
5, I summarize and discuss the generalization to non-Abelian
theories. I also speculate on a possible application in strong
interaction physics.

\section{Chiral Poincar\'{e} transformations}

The model we consider is a Dirac fermion $\psi$ with mass $m$ coupled
to an external vector field $A_\mu$. The simplicity of this model
allows us to discuss the main ideas without the complications of
non-Abelian structure or couplings to other fields. The Lagrangian is
\begin{eqnarray}
{\cal L} & = & \bar\psi[i\dslash-\;\fslash\! A-m]\psi.
\end{eqnarray}
$A_\mu$ can be a gauge potential or an arbitrary external source.

We first recall the elementary discussion of phase and Poincar\'{e}
transformations. This is useful for comparison. Since we are
considering an Abelian model, the phase transformations are generated
by the unit operator.  The Poincar\'{e} transformations are generated
by
\begin{eqnarray}
P_\mu & = & i\partial_\mu,
\end{eqnarray}
the generator of translations, and
\begin{eqnarray}
J_{\mu\nu} & = & \half\sigma_{\mu\nu}
+(x_\mu i\partial_\nu-x_\nu i\partial_\mu)
  \;\equiv\; S_{\mu\nu}+L_{\mu\nu},
\end{eqnarray}
the generator of Lorentz transformations.

Our transformations are {\em active}. That is, they transform the
physical system, the fermion field, in contrast to {\em passive}
transformations where the coordinate system is transformed.  This is
essential for the chiral transformations. Thus no transformation law
is assigned to the $A_\mu$-field.

The transformations are:
\begin{enumerate}

\item Phase transformations:
\begin{eqnarray}
\psi(x) & \to & e^{i\alpha}\psi(x),
  \hspace{2em} \bar\psi(x)
  \; \to \; \bar\psi(x)e^{-i\alpha}
\end{eqnarray}
$\alpha$ is a dimensionless constant. The Lagrangian is invariant.

\item Translations:
\begin{eqnarray}
\label{psi-trans}
\psi(x) & \to & e^{ia_\mu P^\mu}\psi(x),
  \hspace{2em} \bar\psi(x)
  \; \to \; \bar\psi(x)e^{-ia_\mu P^\mu}
\end{eqnarray}
$a_\mu$ is a constant vector with the dimension of length. For $a_\mu$
infinitesimal, ${\cal L}$ transforms into
\begin{eqnarray}
\label{L-trans}
{\cal L} & \to & \bar\psi[i\dslash
-\;\fslash\! A-a_\nu\partial^\nu\!\fslash\! A-m]\psi.
\end{eqnarray}
If we assign the transformation law
\begin{eqnarray}
\label{A-trans}
A_\mu & \to & A_\mu-a_\nu\partial^\nu\! A_\mu
\end{eqnarray}
to $A_\mu$ to accompany (\ref{psi-trans}), we will have that ${\cal
L}$ is invariant. The infinitesimal transformation (\ref{A-trans}) is
of course identical to that found from a translation of the coordinate
system.

\item Lorentz transformations:
\begin{eqnarray}
\psi(s) & \to & e^{i\frac{1}{2}\omega_{\mu\nu}
J^{\mu\nu}}\psi(x),
  \hspace{2em} \bar\psi(x)
  \;\to\; \bar\psi e^{-i\frac{1}{2}\omega_{\mu\nu}J^{\mu\nu}}
\end{eqnarray}
$\omega_{\mu\nu}$ is a dimensionless constant antisymmetric
tensor. For infinitesimal $\omega_{\mu\nu}$ we have
\begin{eqnarray}
{\cal L} & \to & \bar\psi[i\dslash-\;\fslash\! A
-\omega^{\mu\nu}A_\nu\gamma_\mu-\half\omega^{\mu\nu}
(x_\mu\partial_\nu\fslash\! A
-x_\nu\partial_\mu\fslash\! A)-m]\psi
\end{eqnarray}
from which we can construct the transformation rule
\begin{eqnarray}
\nonumber
A_\mu & \to & A_\mu+i\half\omega^{\rho\sigma}
(J_{\rho\sigma})_{\mu\nu}A^\nu \\
(J_{\rho\sigma})_{\mu\nu} & \equiv & i(g_{\rho\mu}g_{\sigma\nu}
-g_{\sigma\mu}g_{\rho\nu})
+(x_\rho i\partial_\sigma-x_\sigma i\partial_\rho)g_{\mu\nu}
\end{eqnarray}
which leads to invariance.
\end{enumerate}
The chiral versions of these are then the following.
\begin{enumerate}

\item Chiral phase transformations:
\begin{eqnarray}
\psi(x) & \to & e^{i\beta\gamma_5}\psi(x),
  \hspace{2em} \bar\psi(x)
  \;\to\; \bar\psi(x)e^{i\beta\gamma_5}
\end{eqnarray}
$\beta$ is a dimensionless pseudoscalar constant. These are the usual
ones, and lead to the Lagrangian
\begin{eqnarray}
\label{wave-phase}
{\cal L} & \to \bar\psi[i\dslash-\;\fslash\! A-m
-2im\beta\gamma_5]\psi.
\end{eqnarray}

\item Chiral translations:
\begin{eqnarray}
\psi(x) & \to & e^{ib_\mu P^\mu\gamma_5}\psi(x),
  \hspace{2em} \bar\psi(x)
  \;\to\; \bar\psi(x)e^{ib_\mu P^\mu\gamma_5}
\end{eqnarray}
$b_\mu$ is a constant axial vector with the dimension of length. This
leads to
\begin{eqnarray}
{\cal L} & \to & \bar\psi[i\dslash-\;\fslash\! A
-b_\mu\partial^\mu\!\fslash\! A\gamma_5
-m+2m\gamma_5b_\mu\partial^\mu]\psi.
\end{eqnarray}

\item Chiral Lorentz transformations:
\begin{eqnarray}
\psi(x) & \to & e^{i\frac{1}{2}\phi_{\mu\nu}
J^{\mu\nu}\gamma_5}\psi(x),
  \hspace{2em} \bar\psi(x)
  \; \to \; \bar\psi(x)e^{i\frac{1}{2}\phi_{\mu\nu}
J^{\mu\nu}\gamma_5}
\end{eqnarray}
$\phi_{\mu\nu}$ is a ``dual'' tensor (i.e.\ it has properties like
$\tilde F_{\mu\nu}$ under $C$, $P$ and $T$) and is dimensionless. This
leads to
\begin{eqnarray}
\label{wave-lor}
{\cal L} & \to & \bar\psi[i\dslash-\;\fslash\! A
-\phi^{\mu\nu}A_\nu\gamma_\mu\gamma_5-\half\phi^{\mu\nu}
(x_\mu\partial_\nu\fslash\! A-x_\nu\partial_\mu\fslash\! A)\gamma_5
-m-im\phi_{\mu\nu}J^{\mu\nu}\gamma_5]\psi.
\end{eqnarray}
\end{enumerate}

It is now impossible to find transformation rules for $A_\mu$ which
restores invariance. Chiral transformations includes $\gamma_5$ in the
infinitesimal generators, and the left- and right-handed parts of the
fermion are transformed in an opposite sense, as can be seen from the
representation where $\gamma_5$ is diagonal:
\begin{eqnarray}
\gamma_5 & = & \left( \begin{array}{rr} I & 0 \\
  0 & -I \end{array} \right).
\end{eqnarray}
In this case there are no interpretations in terms of transformations
of the coordinate system, but as long as they are regarded as active
the chiral Poincar\'{e} transformations make just as much sense as the
non-chiral ones. Note also that new derivative operators have
been generated in the Dirac operator.

\section{Proper time representation and Poincar\'{e} invariance}

I now turn to the the quantum theory and the calculation of anomalies,
or in other words, Jacobians. I use proper time regularization of the
fermionic determinant \cite{ball}, which I will discuss in some detail
even if it is a well known scheme. The reasons for this are, first,
that I work in Minkowski space and that there are differences from the
usual Euclidean formulation. Second, I need to demonstrate that the
scheme is both phase and Poincar\'{e} invariant. And third, compared
to the calculation of the ABJ anomaly, there are additional features
coming from the $\sigma_{\mu\nu}$ and the derivative operators in the
generators. The calculations in this section are essentially model
independent.

In Minkowski space, ambiguities are resolved by adding a small
positive imaginary part to the Dirac operator:
\begin{eqnarray}
\label{operator}
{\cal L} & = & \bar\psi(D+i\epsilon)\psi, \hspace{2em}
D \;=\; i\dslash-\;\fslash\! A-m.
\end{eqnarray}
This is the usual $\epsilon$-prescription which will render all
momentum integrals well defined, and which gives the fermion mass a
small negative imaginary part, i.e.\ the fermion mass is $m_c\equiv
m-i\epsilon$.

The quantum theory is expressed by the path integral $Z$, which
defines the fermionic determinant,
\begin{eqnarray}
Z & = & \int{\cal D}\psi{\cal D}\bar\psi e^{i\int d^4x\bar\psi D\psi}
  \;\equiv\; \mbox{Det}D,
\end{eqnarray}
where $\psi$ and $\bar\psi$ are the independent quantities, and the
effective action $W$ ($Z\equiv e^{iW}$) is given by
\begin{eqnarray}
W & = & -i\Tr\ln D.
\end{eqnarray}
These quantities are formal and will become well defined when we
regularize the theory.

I will calculate the Jacobian of a local infinitesimal change of path
integration variables.  Let us collectively denote the phase and
Poincar\'{e} transformations by
\begin{eqnarray}
\psi & \to & e^{iA}\psi,
  \hspace{2em} \bar\psi \;\to\; \bar\psi e^{-iA},
\end{eqnarray}
where
\begin{eqnarray}
A & = & \alpha, \hspace{1em} \half(a_\mu P^\mu+P^\mu a_\mu),
  \hspace{1em} \mbox{or} \hspace{1em} \half(\half\omega_{\mu\nu}
J^{\mu\nu}+\half J_{\mu\nu}\omega^{\mu\nu}).
\end{eqnarray}
The parameters are now local, $\alpha=\alpha(x)$, etc. We use
symmetric products for the local translations and Lorentz
transformations, since these are Dirac hermitian. Similarly, their
chiral counterparts are
\begin{eqnarray}
\psi & \to & e^{iB\gamma_5}\psi,
  \hspace{2em} \bar\psi \;\to\; \bar\psi e^{iB\gamma_5},
\end{eqnarray}
with
\begin{eqnarray}
B & = & \beta, \hspace{1em} \half(b_\mu P^\mu+P^\mu b_\mu),
  \hspace{1em} \mbox{or} \hspace{1em} \half(\half\phi_{\mu\nu}
J^{\mu\nu}+\half J_{\mu\nu}\phi^{\mu\nu}).
\end{eqnarray}
These transformations induce a change in the Dirac operator:
\begin{eqnarray}
\label{variation}
\nonumber
D & \to & e^{-iA+iB\gamma_5}De^{iA+iB\gamma_5} \\
\nonumber
  & = & D+i(DA-AD)+i(DB\gamma_5+B\gamma_5D) \\
  & \equiv & D+\delta D,
\end{eqnarray}
which in turn induces a change in $W$:
\begin{eqnarray}
\delta W & = & -i\Tr\delta D\frac{1}{D}.
\end{eqnarray}
The Jacobian $J$ is then determined by the requirement that the path
integral $Z$ is unchanged by a change of variables:
\begin{eqnarray}
Z & = & Je^{iW+i\delta W} \;\equiv\; e^{iS_J}e^{iW+i\delta W}
  \;=\; e^{iW},
\end{eqnarray}
where we have defined the action $S_J\equiv-i\ln J$. Therefore $S_J$
can be found by
\begin{eqnarray}
S_J & = & -\delta W \;=\; i\Tr\delta D\frac{1}{D}.
\end{eqnarray}
This is the quantity we are interested in, and which contains the
anomalies.

We now introduce an operator $\tilde D$ and a proper time integral,
thereby defining the formal trace:
\begin{eqnarray}
\label{proper}
\nonumber
S_J & = & i\Tr\delta D\frac{1}{D} \\
\nonumber
  & = & i\Tr\delta D\tilde D\frac{1}{D\tilde D} \\
  & = & \int_{1/\Lambda^2}^\infty ds\Tr\delta D\tilde D
e^{is(D\tilde D+i\epsilon)}
\end{eqnarray}
The operator $\tilde D$ is a priori arbitrary, except that it must be
chosen to give the right $\epsilon$-prescription, like I have written
here. This will then ensure convergence at the upper integration
limit. For the lower integration limit the cutoff $\Lambda$ is
introduced, which is to be taken to infinity at the end of the
calculation. I will discuss the appropriate choice for $\tilde D$
below. When this choice is made $S_J$ will be regular and well
defined.

We can use the expression for $\delta D$ (eq.\ (\ref{variation})) to
perform the proper time integral and write $S_J$ in a Fujikawa-like
form \cite{fujikawa} (see also \cite{petersen}):
\begin{eqnarray}
\label{fujikawa}
\nonumber
S_J & = & \int_{1/\Lambda^2}^\infty ds
\Tr\, i(DA-AD+DB\gamma_5+B\gamma_5D)\tilde De^{isD\tilde D} \\
  & = & -\Tr\, A\left(e^{i\tilde DD/\Lambda^2}
-e^{iD\tilde D/\Lambda^2}\right)
-\Tr\, B\gamma_5\left(e^{i\tilde DD/\Lambda^2}
+e^{iD\tilde D/\Lambda^2}\right).
\end{eqnarray}
Here I have used the cyclicity of the trace, the identity $\tilde
De^{isD\tilde D}D = \tilde DDe^{is\tilde DD}$, and the fact that only
the lower limit of the integral contributes due to the implicit
presence of the $\epsilon$.

We must now make an appropriate choice for $\tilde D$, one which
preserves phase and Poincar\'{e} invariance and leads to the right
$\epsilon$-prescription. In principle $\tilde D$ can still be
completely unrelated to $D$; for instance we can choose the unit
operator, $\tilde D=I$. But the choice I will adopt, to be justified
in a moment, is
\begin{eqnarray}
\label{D-tilde}
\tilde D & = & (i\gamma_5)D(i\gamma_5).
\end{eqnarray}
For our QED-like theory we get $\tilde D=i\dslash-\fslash A+m$, with
{\em plus} in front of the mass, and
\begin{eqnarray}
\tilde DD & = & D\tilde D
  \;=\; -D_\mu D^\mu-\half\sigma_{\mu\nu}F^{\mu\nu}-m^2,
  \hspace{2em} D_\mu \;=\; \partial_\mu+iA_\mu,
\end{eqnarray}
the familiar ``square'' of the Dirac operator. $i\epsilon$ is
implied. Using the cyclicity of the trace and the fact that $\gamma_5$
commutes with $A$, we have
\begin{eqnarray}
\nonumber
S_J & = & -2\Tr\, B\gamma_5e^{i\tilde DD/\Lambda^2}.
\end{eqnarray}
The terms proportional to $A$ in eq.\ (\ref{fujikawa}) have thus
cancelled out. The vector current, energy-momentum tensor and angular
momentum tensor are then conserved, and both phase and Poincar\'{e}
invariance are intact.

I have two reasons for the choice (\ref{D-tilde}) of $\tilde D$.
First, I want to relate it to $D$ such that $\tilde DD$ and $D\tilde
D$ are second order derivative operators.  This simplifies later
calculations.  Second, it automatically respects both phase and
Poincar\'{e} invariance. Furthermore, in a non-Abelian model with an
external vector $V_\mu^a$ and axial vector $A_\mu^a$, this choice
gives the Bardeen anomaly -- the Wess--Zumino consistent anomaly.

\section{Jacobians for the chiral transformations}

For the calculation of the Jacobians, it is necessary to consider each
form of $B$ separately. As I have already mentioned, there are two
further ingredients in the calculation of the chiral Poincar\'{e}
Jacobians compared to the chiral phase case, namely the occurrence of
$\sigma_{\mu\nu}$ in $J_{\mu\nu}$ and of derivative operators in
$P_\mu$ and $J_{\mu\nu}$.

Let us write
\begin{eqnarray}
\tilde DD & = & -D_\mu D^\mu-Y,
  \hspace{2em} Y \;=\; \half\sigma_{\mu\nu}F^{\mu\nu}+m^2.
\end{eqnarray}
It is useful first to recall the familiar calculation of the ABJ
anomaly \cite{fujikawa}
\begin{eqnarray}
S_\beta & = & -2\Tr\,\beta\gamma_5e^{i\tilde DD/\Lambda^2}
  \;\equiv\; \int d^4x{\cal L}_\beta.
\end{eqnarray}
${\cal L}_\beta$ is given by
\begin{eqnarray}
\nonumber
{\cal L}_\beta & = & -2\int\frac{d^4k}{(2\pi)^4}e^{ikx}
\tr\beta\gamma_5e^{i(-D^2-Y)/\Lambda^2}e^{-ikx} \\
\nonumber
  & = & -2\int\frac{d^4k}{(2\pi)^4}\tr\beta\gamma_5
e^{i(k^2+2ikD-D^2-Y)/\Lambda^2} \\
  & = & -2\Lambda^4\int\frac{d^4k}{(2\pi)^4}e^{ik^2}
\tr\beta\gamma_5\exp i\left(\frac{2ikD}{\Lambda}
-\frac{D^2+Y}{\Lambda^2}\right),
\end{eqnarray}
where I have scaled $k_\mu\to\Lambda k_\mu$. Expanding the exponential
and keeping only terms that are not suppressed by the cutoff
$\Lambda$, we get
\begin{eqnarray}
\nonumber
{\cal L}_\beta & = & -2\int\frac{d^4k}{(2\pi)^4}e^{ik^2}
\tr\gamma_5\sigma_{\mu\nu}\sigma_{\rho\sigma}\beta
(-\mbox{$\frac{1}{8}$}F^{\mu\nu}F^{\rho\sigma}) \\
  & = & \frac{1}{8\pi^2}\beta F\tilde F.
\end{eqnarray}

In comparison we have for chiral translations
\begin{eqnarray}
\nonumber
{\cal L}_b & = & -2\int\frac{d^4k}{(2\pi)^4}e^{ikx}
\tr\left(\half b_\mu i\partial^\mu
+\half i\partial^\mu b_\mu\right)\gamma_5
e^{i\tilde DD/\Lambda^2}e^{-ikx} \\
\nonumber
  & = & -2\int\frac{d^4k}{(2\pi)^4}
\tr\left[-\half i(\partial_\mu b^\mu)+b_\mu k^\mu\right]\gamma_5
e^{i(k^2+2ikD-D^2-Y)/\Lambda^2} \\
  & \equiv & {\cal L}_b^{(1)}+{\cal L}_b^{(2)}
\end{eqnarray}
I have used partial integration to bring the first term into the form
of the ABJ anomaly, for which we get
\begin{eqnarray}
{\cal L}_b^{(1)} & = & \frac{1}{8\pi^2}
(-i\half\partial_\mu b^\mu)F\tilde F.
\end{eqnarray}
The second term, after scaling, is
\begin{eqnarray}
\nonumber
{\cal L}_b^{(2)} & = & -2\Lambda^5\int\frac{d^4k}{(2\pi)^4}
e^{ik^2}\tr b_\mu k^\mu\gamma_5\exp i\left(\frac{2ikD}{\Lambda}
-\frac{D^2+Y}{\Lambda^2}\right) \\
\nonumber
  & = & -2\int\frac{d^4k}{(2\pi)^4}e^{ik^2}k_\mu k_\nu
\tr\gamma_5\sigma_{\rho\sigma}\sigma_{\alpha\beta} \\
\nonumber
  & & \mbox{} \hspace{4em} \times b^\mu\left(\mbox{$\frac{1}{12}$}
D^\nu F^{\rho\sigma}F^{\alpha\beta}
+\mbox{$\frac{1}{12}$}F^{\rho\sigma}D^\nu F^{\alpha\beta}
+\mbox{$\frac{1}{12}$}F^{\rho\sigma}F^{\alpha\beta}D^\nu\right) \\
  & = & \frac{1}{16\pi^2}(i\partial_\mu b^\mu)F\tilde F
+\frac{1}{8\pi^2}b_\mu A^\mu F\tilde F.
\end{eqnarray}
Summing the two contributions, we get
\begin{eqnarray}
\label{chi-trans}
{\cal L}_b & = & \frac{1}{8\pi^2}b_\mu A^\mu F\tilde F
\end{eqnarray}
This expression is not invariant under the gauge transformation
$A_\mu\to A_\mu-\partial_\mu\alpha$ (when $A_\mu$ is a gauge
potential). This is slightly surprising, since we were careful about
gauge invariance when we regularized our theory. I will discuss this
point in the next section.

For the chiral Lorentz transformations, there are again two terms
which are handled essentially in this way, but in addition there is
now a third term, coming from the $\half\sigma_{\mu\nu}\gamma_5$ in
$J_{\mu\nu}\gamma_5$, which requires special attention. This time the
trace leads to only one factor of $F_{\mu\nu}$, and the term computes
to
\begin{eqnarray}
\nonumber
{\cal L}_\phi^{(3)} & \equiv & -2\int\frac{d^4k}{(2\pi)^4}e^{ikx}
\tr(\half\phi_{\mu\nu}\half\sigma^{\mu\nu}\gamma_5)
e^{i\tilde DD/\Lambda^2}e^{-ikx} \\
  & = & \frac{1}{48\pi^2}\phi_{\mu\nu}(\square+6m^2)\tilde F^{\mu\nu}.
\end{eqnarray}
The complete chiral Lorentz anomaly is given by
\begin{eqnarray}
\label{chi-lor}
{\cal L}_\phi & = & \frac{1}{48\pi^2}\phi_{\mu\nu}
(\square+6m^2)\tilde F^{\mu\nu}
+\frac{1}{8\pi^2}\half\phi_{\mu\nu}(x^\mu A^\nu
-x^\nu A^\mu)F\tilde F
\end{eqnarray}
In addition to being gauge non-invariant, this is also not manifestly
Poincar\'{e} invariant. However, this is an artifact of writing the
anomaly in terms of a Lagrangian since we integrate this expression
over spacetime to find the action: $S_\phi=\int d^4x{\cal
L}_\phi$. The integration then picks out only the Poincar\'{e}
invariant part of ${\cal L}_\phi$, and physics does not depend on the
coordinate system.

\section{Discussion}

To summarize, I used proper time regularization in Minkowski space to
regularize a theory of a massive Dirac fermion coupled to an external
vector field. When the regularization is chosen to respect phase
rotation and Poincar\'{e} invariance, this lead to a nontrivial
transformation of the fermionic measure under chiral translations and
Lorentz transformations in the path integral. I emphasize that the
chiral Poincar\'{e} transformations are ``anomalous transformations''
and not ``anomalous symmetries''. The ``anomalies'' depend on external
sources and the mass of the fermion, which at the same time break the
would-be symmetry explicitly.

The existence of a Jacobian for infinitesimal chiral Poincar\'{e}
transformations implies of course that finite transformations give
rise to a Jacobian as well, which is what we want for the four
dimensional bosonization procedure. Such a finite Jacobian is a
complicated nonlinear quantity with infinitely many derivatives, and
is very hard to calculate. Furthermore, if the bosonic fields
$\theta$, $\phi_{\mu\nu}$ and $b_\mu$ are relevant degrees of freedom
for bosonization, and if ${\cal L}_J$ is to be the bosonized
Lagrangian it is necessary to make them decouple from the
fermion. Work on this problem is in progress \cite{thomassen}.

Let us return to the problem of the gauge non-invariance of the
anomalies, eqs.\ (\ref{chi-trans}) and (\ref{chi-lor}). Since we were
careful with choosing a gauge invariant regularization of the theory,
the full path integral cannot depend on the part of $A_\mu$ that is a
gradient. That is, if we make the Hodge decomposition
\begin{eqnarray}
A_\mu & = & \partial_\mu\eta+\half\epsilon_{\mu\nu\rho\sigma}
\partial^\nu\xi^{\rho\sigma},
\end{eqnarray}
it does not depend on $\partial_\mu\eta$. We are therefore allowed to
make the replacement
\begin{eqnarray}
A_\mu & \to & \Pi_\mu^{\;\;\nu} A_\nu\equiv\left(g_\mu^{\;\;\nu}
-\frac{\partial_\mu\partial^\nu}{\square}\right)A_\nu
  \;=\; \frac{1}{\square}\partial^\nu F_{\nu\mu}
\end{eqnarray}
in all expressions, where $\Pi_{\mu\nu}$ is a projection operator
which removes the gradient from $A_\mu$. Due to our gauge invariant
regularization, this can be done without loss of generalization.

Let me also remark on our choice of regularization operator $\tilde
D$.  Another reasonable choice would be $\tilde D=-CD^TC^{-1}$, where
$C$ is the charge conjugation matrix and the transposition is with
respect to the Dirac matrix structure. This is the ``charge
conjugate'' of $D$. It gives identical results for our QED-like
theory, but in the non-Abelian case with a vector and an axial vector
it leads to the covariant chiral anomaly instead of the Bardeen
anomaly, and thus the vector current is not conserved in this
case. However, other terms proportional to $A$ in eq.\
(\ref{fujikawa}) are non-zero as well. The energy-momentum tensor and
angular momentum tensor are then also not conserved. In other words,
Poincar\'{e} invariance is lost!

It is possible to generalize the discussion to non-Abelian
fermions. Let us call the internal quantum number `flavor'. First of
all, since we have taken the active point of view for our
transformations, it is possible to ``translate'' or ``Lorentz
transform'' each flavor of fermion separately. These transformations
are generated by $t^aP_\mu$ and $t^aJ_{\mu\nu}$, where $t^a$ are
generators of rotations in flavor space. Their chiral counterparts are
generated by $t^aP_\mu\gamma_5$ and $t^aJ_{\mu\nu}\gamma_5$. All of
these transformations are naive symmetries of the fermionic measure,
but none of them are symmetries of the Lagrangian, except when the
theory under consideration is trivial. They are also not the
generators of a group, since their algebra does not close. However,
the chiral set of generators gives rise to nontrivial Jacobians, and
for this reason these may still, together with chiral phase rotations,
play a role in four dimensional non-Abelian bosonization.

Finally, I will speculate on a possible application of flavored chiral
Poincar\'{e} transformations in strong interaction physics. Recall
that hadrons can be classified, with reasonable success, in terms of
the group $SU(6)$, which is assumed to contain
$SU(2)_{\mathrm{spin}}\times SU(3)_{\mathrm{flavor}}$ \cite{donoghue}.
According to the $SU(6)$ scheme, the $\rho$'s enter in the same
multiplet as the $\pi$'s. But the $\pi$'s can be understood within QCD
as Goldstone bosons of spontaneously broken chiral symmetry, which
raises the question of whether the $\rho$'s could be Goldstone bosons
as well. This question was discussed in a paper by Caldi and Pagels
\cite{caldi}. However, shortly after the discovery of $SU(6)$ in the
60's, it became clear that a symmetry group that mixes internal and
spacetime degrees of freedom in any but a trivial way does not exist
for a relativistic theory (see the lecture note and reprint volume by
Dyson \cite{dyson}). This means that the $\rho$'s cannot correspond to
the broken generators of any symmetry group. Caldi and Pagels
suggested that the $\rho$'s were nevertheless ``dormant'' Goldstone
bosons in a certain static limit.

I claim that the $\rho$'s correspond to chiral Lorentz
transformations, and that these are in some sense ``broken''. This is
not in conflict with the no-go theorems of the 60's \cite{dyson}
because the chiral Lorentz transformations are neither symmetries nor
a group. If we consider eqs.\ (\ref{wave-phase}) and (\ref{wave-lor})
in flavor space, we can read off from the mass terms the quark
wavefunctions that correspond to the transformation parameters. They
are
\begin{eqnarray}
\nonumber
\beta^a & \sim & \bar qit^a\gamma_5q, \\
\phi_{\mu\nu}^a & \sim & \bar qit^a\half J_{\mu\nu}\gamma_5q
  \;=\; \bar qit^a\mbox{$\frac{1}{4}$}\sigma_{\mu\nu}\gamma_5q
+\bar qit^a\half L_{\mu\nu}\gamma_5q.
\end{eqnarray}
The latter implies
\begin{eqnarray}
\tilde\phi_{\mu\nu}^a & \sim & \bar qt^a\mbox{$\frac{1}{4}$}
\sigma_{\mu\nu}q + \mbox{orbital part}.
\end{eqnarray}
The ``electric'' part of the spin part of $\tilde\phi_{\mu\nu}$ is the
same wavefunction for the $\rho$-meson as that suggested in ref.\
\cite{caldi}. Thus, if the field $\beta^a$ describes the pseudoscalar
octet $(\pi,K,\eta)$, and the field $\phi_{\mu\nu}^a$ describes the
vector mesons $(\rho,K^*,\phi)$ and $\omega$ (for $t^0\equiv I$), then
they correspond to quark wavefunctions with the correct quantum
numbers. Perhaps it is possible to test this idea by the methods of
chiral perturbation theory.

\noindent
\paragraph{Acknowledgments} I thank M.\ Faber and A.N.\ Ivanov for
comments and discussions, and P.H.\ Damgaard for calling my attention
to some of the references.

\end{document}